\documentclass[11pt,bezier]{article}
\usepackage{amsmath,amssymb,amsfonts,euscript}
\usepackage[usenames]{color}
\newpage

\newcommand{\be}{\begin{equation}}
\newcommand{\ee}{\end{equation}}

\title{\bf\Large Testing Exponentiality Based on R\'enyi Entropy With Progressively Type-II Censored Data }
\vspace{-1cm}
\author
{Akram Kohansal, Saeid Rezakhah \footnote{Faculty of Mathematics and Computer Science, Amirkabir University of Technology,
Tehran, Iran. Email: ak\_kohansal@aut.ac.ir, rezakhah@aut.ac.ir }}
\begin{document}
\maketitle

\begin{abstract}
We express the joint R\'enyi entropy of progressively censored order statistics in terms of an incomplete integral of the hazard function, and provide a simple estimate of the joint R\'enyi entropy of progressively Type-II censored data. Then we establish a goodness of fit test statistic based on the R\'enyi Kullback-Leibler information with the progressively Type-II censored data, and compare its performance with the leading test statistic. A Monte Carlo simulation study shows that the proposed test statistic shows better powers than the leading test statistic against the alternatives with monotone increasing, monotone decreasing and nonmonotone hazard functions.
\end{abstract}
\section{Introduction}
\par
Suppose that $X$ is a random variable having an absolutely continuous distribution with density function $f(x)$. The R\'enyi entropy of order $\alpha$ is defined as
\begin{equation}\label{ha}
H^{\alpha}(f)=\frac{1}{1-\alpha}\log\int_{-\infty}^{\infty}\{f(x)\}^{\alpha}\mathrm{d}x~~~~\alpha>0,\alpha\neq1,
\end{equation}
and Shannon entropy as $H(f)=\lim_{\alpha\rightarrow1} H^{\alpha}(f)=-\int_{-\infty}^{\infty}f(x)\log f(x)\mathrm{d}x$ provided these integrals exist. The R\'enyi entropy was axiomatized by R\'enyi \cite{28 Pasha} and was modified by some authors \cite{13 Pasha}. R\'enyi entropy has a number of applications in different fields such as statistics \cite{1 Pasha}, \cite{14 Pasha}, \cite{15 Pasha}, \cite{19 Pasha}, \cite{33 Pasha}, biomedical engineering \cite{18 Pasha}, statistical mechanics \cite{8 Pasha}, \cite{22 Pasha}, economics \cite{3 Pasha}, stochastic processes \cite{12 Pasha}, \cite{17 Pasha}, \cite{Pasha}, and some other areas \cite{4 Pasha}, \cite{9 Pasha}, \cite{24 Pasha}, \cite{25 Pasha}, \cite{27 Pasha}, \cite{29 Pasha}.
\par
The R\'enyi Kullback-Leibler (R-KL) information in favor of $f(x)$ against $g(x)$ is defined as
\begin{equation*}
I^\alpha(f;g)=\frac{1}{\alpha-1}\log\left(\int_{-\infty}^{\infty}\frac{\left\{f(x)\right\}^\alpha}{\left\{g(x)\right\}^{\alpha-1}}\mathrm{d}x\right)~~~~\alpha>0,\alpha\neq1,
\end{equation*}
and $I(f;g)=\lim_{\alpha\rightarrow1} I^{\alpha}(f;g)=-\int_{-\infty}^{\infty}f(x)\log \frac{f(x)}{g(x)}\mathrm{d}x$ is the Kullback-Leibler (KL) information if both integrals exist. Because $I^\alpha(f;g)$ has the property that $I^\alpha(f;g)\geq0$, and the equality holds if and only if $f=g$, the estimate of the R-KL information can be consider as a goodness-of-fit test statistic.
\par
From the KL information point of view, the problem of goodness of fit test based on complete sample studied by some authors including \cite{1 Bala}, \cite{8 Bala} and \cite{18 Bala}. Also based on censored sample, \cite{15 Bala} presented a new test statistic with Type-II censored data and \cite{Arghami} modified the previous test statistic. In the case of progressively censored data, \cite{3 My paper} studied the testing exponentiality based on KL information with progressively Type-II censored data and goodness of fit test based on KL information for progressively Type-II censored data can be found in \cite{13 My paper}.
\par
Now we are to study the performance of test statistics based on R-KL information for progressively Type-II censored data. We provide a new test statistic based on R-KL information and compare the power of this test statistic with the performance of Shannon KL information, \cite{3 My paper}. We show through simulation studies that the proposed goodness-of-fit test is more powerful than the test statistic presented in \cite{3 My paper} for different choices of sample sizes and progressive censoring schemes in nearly all cases. However, we can instantly find that this work is not so straight, as we need to estimate the joint R\'enyi entropy of progressively Type-II censored data, via a multi-dimensional integral.
\par
In many life-testing and reliability studies, the complete information may not be available on failure times of all experimental units. There are also situations where the removal of units prior to failure is pre-planned in order to reduce the cost and time associated with testing. For these, and other
reasons, progressive censoring has been considered, see \cite{12 Bala} and \cite{3 My paper}. The conventional Type-I, and Type-II censoring schemes are not flexible enough to allow removal of units at points other than the terminal points of the experiment. So we consider the progressively Type-II censoring as a more general censoring scheme.
\par
The progressive Type-II censoring arises in a life-testing experiment as follows. Suppose $n$ units are placed on test. At the time of the first failure, $R_1$ units are randomly removed from the $n-1$ surviving units. Then at the time of the second failure, $R_2$ units are randomly removed from the $n-R_1-2$ surviving units and so on. Finally after the $m$th failure, all remaining $R_m$ units are removed. Thus, we observe $m$ complete failures and $R_1+R_2+\cdots+R_m$ items are progressively censored from the $n$ units under the test, and so $n=m+(R_1+R_2+\cdots+R_m)$. The vector $R=(R_1,\cdots,R_m)$ is called the progressive censoring scheme and it is fixed prior to the study. A schematic illustration of the progressively Type-II censoring scheme is presented as:

{\vspace{1.5cm}
\hspace{1cm}
\small
\begin{picture}(10,30)(10,15)
\put(25,25){\vector(1,0){240}}
\put(35,25){\vector(1,1){20}}
\put(70,25){\vector(1,1){20}}
\put(170,25){\vector(1,1){20}}
\put(220,25){\vector(1,1){20}}
\put(25,15){\scriptsize $X_{1:m:n}$}
\put(50,50){\scriptsize $R_1$}
\put(60,15){\scriptsize $X_{2:m:n}$}
\put(85,50){\scriptsize $R_2$}
\put(145,15){\scriptsize $X_{m-1:m:n:n}$}
\put(180,50){\scriptsize $R_{m-1}$}
\put(205,15){\scriptsize $X_{m:m:n}$}
\put(230,50){\scriptsize $R_{m}$}
\put(112,20){\scriptsize $...$}
\put(35,25){\circle*{1}}
\put(70,25){\circle*{1}}
\put(170,25){\circle*{1}}
\put(15,-2){\scriptsize \label{fig1} A schematic presentation for progressively Type-II censored scheme.}
\end{picture}
\vspace{0.5in}}

If $R=(0,\cdots,0)$, no withdrawals are made and it corresponds to the complete sample situation in which case the usual order statistics will be obtained. If $R=(0,\cdots,0,n-m)$, we obtain the conventional Type-II right censoring scheme. We will denote the progressively Type-II censored samples as $X_{1:m:n}<X_{2:m:n}<\cdots<X_{m:m:n}$. Progressive censoring scheme, methodology, goodness of fit, estimation have been studied in \cite{2 Bala}, \cite{4 Bala}, \cite{5 Bala}, \cite{13 Bala} and \cite{19 Bala}. A book-length account is available in \cite{3 Bala}.
\par
The joint R\'enyi entropy of $X_{1:m:n},\cdots,X_{m:m:n}$ is simply defined to be
$$\hspace{-.5in}H^\alpha_{1\cdots m:m:n}=\frac{1}{1-\alpha}\log \int_{-\infty}^{\infty}\cdots\int_{-\infty}^{x_{2:m:n}}\left\{f_{X_{1:m:n},\cdots,X_{m:m:n}}(x_1,\cdots,x_m)\right\}^\alpha$$$$\hspace{2.5in}\mathrm{d}x_{1}\cdots \mathrm{d}x_{m}~~~~\alpha>0,\alpha\neq1.$$
In the above formula, $f_{X_{1:m:n},\cdots,X_{m:m:n}}(x_1,\cdots,x_m)$ is the joint p.d.f. of all $m$ progressively Type-II right censored order statistics $(X_{1:m:n},\cdots,X_{m:m:n})$, given in \cite{3 Bala} as
\begin{equation*}
f_{X_{1:m:n},\cdots,X_{m:m:n}}(x_1,\cdots,x_m)=c\prod_{i=1}^{m}f(x_i)\{1-F(x_i)\}^{R_i}~~~~x_1<\cdots<x_m,
\end{equation*}
where
\begin{equation*}
c=n(n-R_1-1)\cdots(n-R_1-R_2-\cdots-R_m+1).
\end{equation*}
The rest of this paper is arranged as follows: In Section II, we first present $H^\alpha_{1\cdots m:m:n}$ as a single-integral in terms of the hazard function, $h(x)$, as
\begin{equation}\label{H}
H^\alpha_{1\cdots m:m:n}=-\log c+\frac{1}{1-\alpha}\sum_{j=1}^{m}\log\int_{-\infty}^{\infty} f_{X_{j:m:n}}(x)\{h(x)\}^{\alpha-1}\left\{1-F_{X_{j:m:n}}(x)\right\}^{\alpha-1}\mathrm{d}x,
\end{equation}
We provide an estimate of (\ref{H}), and define the R-KL information for progressively Type-II censored data. In Section 3, we propose a goodness-of-fit test for exponentiality based on this R-KL information, and use Monte Carlo simulations to evaluate the power under different progressive Type-II censoring schemes. We also compared the performance of our test statistic with the test presented by \cite{13 My paper}. Finally, in Section 4, we present an illustrative example.

\section{R\'enyi Entropy of Progressively Censored Data in Terms of Hazard Function }
\subsection{R\'enyi Entropy Representation}
\par
Another expression of (\ref{ha}) is presented in terms of the hazard function as
\begin{equation}\label{3b}
H_{1:1:1}^\alpha=\frac{1}{1-\alpha}\log\int_{-\infty}^{\infty}f(x)\{h(x)\}^{\alpha-1}\{1-F(x)\}^{\alpha-1}\mathrm{d}x.
\end{equation}
We first note that (\ref{3b}) gives a simple expression of $H_{1:m:n}^{\alpha}$ as
\begin{equation}\label{4b}
H_{1:m:n}^{\alpha}=-\log n+\frac{1}{1-\alpha}\log\int_{-\infty}^{\infty}f_{X_{1:m:n}}(x)\{h(x)\}^{\alpha-1}\{1-F(x)\}^{n(\alpha-1)}\mathrm{d}x.
\end{equation}
Theorem 2.1 below states that the multiple integral in $H^\alpha_{1\cdots m:m:n}$ can be simplified to a single integral.
\par
{\em Theorem 2.1:}
\begin{equation*}
H^{\alpha}_{1\cdots m:m:n}=-\log c+\bar{H}^{\alpha}_{1\cdots m:m:n},
\end{equation*}
where
\begin{equation*}
\bar{H}^{\alpha}_{1\cdots m:m:n}=\frac{1}{1-\alpha}\sum_{j=1}^{m}\log\int_{-\infty}^{\infty}f_{X_{j:m:n}}(x)\{h(x)\}^{\alpha-1}\left\{1-F_{X_{j:m:n}}(x)\right\}^{\alpha-1}\mathrm{d}x.
\end{equation*}
{\bf Proof}: See Appendix A.
\\
The function $\bar{H}^{\alpha}_{1\cdots m:m:n}$ in Theorem (2.1) can be expressed in terms of $\log f(x)$ as follows.
\par
{\em Lemma 2.1:}
\begin{equation*}
\bar{H}^{\alpha}_{1\cdots m:m:n}=\frac{1}{1-\alpha}\sum_{j=1}^{m}\log E\left(\int_{o}^{U_{(1:\gamma_{i}-1)}}c_{j-1}\sum_{i=1}^ja_{i,j}\frac{\{\frac{\mathrm{d}F^{-1}(p)}{\mathrm{d}p}\}^{1-\alpha}}{(1-p)^{\alpha-1}}\right.
\end{equation*}
\begin{equation}\label{5b}
\left.\times\left[\sum_{u=0}^{j-1}\binom{n}{u}p^u(1-p)^{n-u}\right]^{\alpha-1}\mathrm{d}p\right),
\end{equation}
where $U_{(1:\gamma_{i}-1)}$ is the first uniform order statistic from a sample of size $\gamma_{i}-1$ and $\gamma_i=m-i+1+\sum_{j=i}^mR_j$ for $1\leq i\leq m$, $c_{j-1}=\prod_{u=1}^j\gamma_u$ for $1\leq j\leq m$, $a_{i,j}=\prod_{u=1}^j1/(\gamma_u-\gamma_i)$ for $1\leq i\leq j\leq m$, and $\gamma_u-\gamma_i=1$ for $u=i$.
\\
{\bf Proof}: See Appendix B.

\subsection{Nonparametric R\'enyi Entropy Estimation}
\par
Now approximating $\bar{H}^{\alpha}_{1\cdots m:m:n}$ in (\ref{5b}) by
\begin{equation*}
\frac{1}{1-\alpha}\sum_{j=1}^{m}\log \int_{o}^{E(U_{(1:\gamma_{i}-1)})}c_{j-1}\sum_{i=1}^ja_{i,j}\frac{\{\frac{\mathrm{d}F^{-1}(p)}{\mathrm{d}p}\}^{1-\alpha}}{(1-p)^{\alpha-1}}
\left[\sum_{u=0}^{j-1}\binom{n}{u}p^u(1-p)^{n-u}\right]^{\alpha-1}\mathrm{d}p,
\end{equation*}
estimating the derivative of $F^{-1}(p)$ by
$$\displaystyle A_{j,w}=\frac{x_{j+w:m:n}-x_{j-w:m:n}}{p_{i+w:m:n}-p_{i-w:m:n}},$$
 where $$p_{i:m:n}=E[F(X_{i:m:n})]=E[U_{i:m:n}],$$ and $$E[U_{i:m:n}]=1-\prod_{j=m-i+1}^{m}\left\{\frac{j+R_{m-j+1}+\dots+R_{m}}{j+1+R_{m-j+1}+\dots+R_{m}}\right\},$$
 and approximating the integral of $\int_{0}^\epsilon g(x)dx$ by a Riemann sum by $\epsilon g(\frac{\epsilon}{2}),$ we obtain an estimate of $\bar{H}^{\alpha}_{1\cdots m:m:n}$ as
\begin{equation*}
H^\alpha(w,n,m)=\frac{1}{1-\alpha}\sum_{j=1}^{m}\log A_{j,w}^{1-\alpha}c_{j-1}\sum_{i=1}^j\frac{a_{i,j}}{\gamma_i(1-\frac{1}{2\gamma_i})^{\alpha-1}}
\end{equation*}
\begin{equation*}
\hspace{2in}\times\left[\sum_{u=0}^{j-1}\binom{n}{u}(\frac{1}{2\gamma_i})^u(1-\frac{1}{2\gamma_i})^{n-u}\right]^{\alpha-1}.
\end{equation*}
Thus, from Theorem 2.1, an estimate of $H^{\alpha}_{1\cdots m:m:n}$ is obtained as
$$H^\alpha_{1\cdots m:m:n}(w,n,m)=-\log c+H^\alpha(w,n,m).$$

\subsection{R\'enyi Kullback-Leibler Information, and Test Statistic}
\par
For a null density function $f^0(x;\theta)$, the R-KL information from progressively Type-II censored data is defined to be
\begin{equation*}
\hspace{-2in}I^\alpha_{1\cdots m:m:n}(f;f^0)=\frac{1}{\alpha-1}\log\int_{-\infty}^{\infty}\cdots\int_{-\infty}^{x_{2:m:n}}
\end{equation*}
\begin{equation*}
\hspace{1.5in}\frac{\{f_{X_{1:m:n},\cdots,X_{m:m:n}}(x_1,\cdots,x_m;\theta)\}^\alpha}
{\{f^0_{X_{1:m:n},\cdots,X_{m:m:n}}(x_1,\cdots,x_m;\theta)\}^{\alpha-1}}\mathrm{d}x_1\cdots \mathrm{d}x_m.
\end{equation*}
Consequently, the R-KL information can be estimated by
\begin{equation}\label{6b}
I^\alpha_{1\cdots m:m:n}(f;f^0)=-H^\alpha_{1\cdots m:m:n}-\sum_{j=1}^m\log f^0(x_j;\theta)
-\sum_{j=1}^mR_j\log(1-F^0(x_j;\theta)).
\end{equation}
Thus, the test statistic based on $I^\alpha_{1\cdots m:m:n}(f;f^0)/n$ is given by
\begin{equation}\label{7b}
T^\alpha(w,n,m)=-\frac{1}{n}H^\alpha(w,n,m)
-\frac{1}{n}\left[\sum_{j=1}^m\log f^0(x_j;\hat{\theta})+\sum_{j=1}^mR_j\log(1-F^0(x_j;\hat{\theta}))\right],
\end{equation}
where $\hat{\theta}$ is an estimator of $\theta$.

\section{Testing Exponentiality Based on The R\'enyi Kullback-Leibler Information}
\subsection{Test Statistic}
\par
Suppose that we are interested in a goodness of fit test for
\begin{equation*}
\left\{
\begin{array}{c}
H_0: f^0(x)=\frac{1}{\theta}\exp(-\frac{x}{\theta}),\\
H_A: f^0(x)\neq\frac{1}{\theta}\exp(-\frac{x}{\theta}),
\end{array}
\right.
\end{equation*}
where $\theta$ is unknown. Then the R-KL information for progressively Type-II censored data can be approximated, in view of (\ref{6b}), with
\begin{equation*}
I^\alpha_{1\cdots m:m:n}(f;f^0)=-H^\alpha_{1\cdots m:m:n}+m\log \theta+\frac{1}{\theta}\sum_{j=1}^m(R_j+1)X_{j:m:n}.
\end{equation*}
If we estimate the unknown parameter $\theta$ by the maximum likelihood estimate, $(\sum_{j=1}^m(R_j+1)X_{j:m:n})/m$, then we have an estimate of $I^\alpha_{1\cdots m:m:n}(f;f^0)/n$ as
\begin{equation}\label{8b}
T^\alpha(w,n,m)=-\frac{1}{n}H^\alpha(w,n,m)+\frac{m}{n}\left[\log\left(\frac{1}{m}\sum_{j=1}^m(R_j+1)X_{j:m:n}\right)+1\right].
\end{equation}
Under the null hypothesis, $T^\alpha(w,n,m)$ for $\alpha$ close to $1$, will be close to $0$, and therefor large values of $T^\alpha(w,n,m)$ will lead to the rejection of $H_0.$

\subsection{Implementation of the Test}
\par
Because the sampling distribution of $T^\alpha(w,n,m)$ is intractable, we determine the percentage points using 10,000 Monte Carlo simulations from an exponential distribution. In determining the window size $w$ which depends on $n$, $m$, $\alpha$ and $\eta$, significance level, we chose the optimal window size $w$ to be one which gives the minimum critical points in the sense of \cite{8 Bala}. Also, we chose $\alpha$ in R-KL information using the trial and error method. However, we find from the simulated percentage points that the optimal window size $w$ varies much according to $m$ rather than $n$, and does not vary much according to $\alpha$, if $\alpha\leq1$. In view of these observations, our recommended values of $w$ for different $m$ are as given in \cite{15 Bala} and our recommended value of $\alpha$, using the trial and error method, is $\alpha=0.4$.
\par
To obtain the critical values, after deciding about the value of $w$ and $\alpha$, simulate the whole procedure by taking the observation from the $E(1)$ distribution, and calculate the value of $T^\alpha(w,n,m)$, for $10,000$ times.

\subsection{Power Results}
\par
There are lots of test statistics for exponentiality concerning uncensored data \cite{2 Park}, \cite{8 Park}, \cite{10 Park}, \cite{11 Park}, \cite{12 Park}, but only some of them can be extended to the censored data. We consider here the test statistic \cite{3 My paper} among them. \cite{3 My paper} proposed the test statistic as
\begin{equation*}
T(w,n,m)=-H(w,n,m)+\frac{m}{n}\left[\log\left(\frac{1}{m}\sum_{i=1}^m(R_i+1)X_{i:m:n}\right)+1\right],
\end{equation*}
where
\begin{equation*}
H(w,n,m)=\frac{1}{n}\sum_{i=1}^m\log\left(\frac{x_{i+w:m:n}-x_{i-w:m:n}}{E(U_{i+w:m:n})-E(U_{i-w:m:n})}\right)
-\left(1-\frac{m}{n}\right)\log\left(1-\frac{m}{n}\right).
\end{equation*}
\par
As the proposed test statistic is related to the hazard function of the distribution, we consider the alternatives according to the type of hazard function as follows.
\begin{itemize}
{\item I) Monotone increasing hazard: Gamma and Weibull (shape parameter 2),}
{\item II) Monotone decreasing hazard: Gamma and Weibull (shape parameter 0.5),}
{\item III) Nonmonotone hazard: Center Beta (shape parameter 0.5), Log normal (shape parameter 1).}
\end{itemize}
We used $10,000$ Monte Carlo simulations for $n=10,\;20$ and $30$ to estimate the power of our proposed test statistic, and the competing test statistic. The simulation results are summarized in Figures 1-3, and Tables 1-3.

\begin{table}
\label{t1}
\caption{ \scriptsize{ Power comparison for different hazard alternatives at 10\% significance level for several progressively censored samples with the sample size is $n=10$}}
\vspace{.05in}
\hspace{-0.5in}
{\begin{tabular}{|c|c|c||c|c||c|c||c|c|c}
\hline
\hline
 & & & \multicolumn{2}{|c||}{\small monotone increasing} & \multicolumn{2}{|c||}{\small monotone decreasing} & \multicolumn{2}{|c|}{\small nonmonotone} \\
 & & & \multicolumn{2}{|c||}{\small hazard alternatives} & \multicolumn{2}{|c||}{\small hazard alternatives} & \multicolumn{2}{|c|}{\small hazard alternatives} \\
\cline{1-9}
{m} & \small  schemes           & statistics &\small Gamma       &\small Weibull     &\small  Gamma       &\small Weibull     & \small  Beta       &\small Log-Normal     \\
    & \small $(R_1,\dots, R_m)$ &  &\small shape 2     &\small shape 2     &\small  shape 0.5   &\small  shape 0.5   & \small  shape 0.5     &\small  shape 1     \\
\hline
5 & 5,0,0,0,0 & $T(w,n,m)$        & 0.406 & 0.690 & 0.020 & 0.026 & 0.039 & 0.219 \\
  &  & $T^\alpha(w,n,m)$ & 0.446 & 0.731 & 0.027 & 0.040 & 0.051 & 0.241 \\
\hline
5 & 0,5,0,0,0 & $T(w,n,m)$        & 0.381 & 0.651 & 0.019 & 0.024 & 0.033 & 0.212 \\
  &  & $T^\alpha(w,n,m)$ & 0.423 & 0.708 & 0.027 & 0.041 & 0.042 & 0.247 \\
\hline
5 & 1,1,1,1,1 & $T(w,n,m)$        & 0.352 & 0.532 & 0.027 & 0.027 & 0.030 & 0.243 \\
  &  & $T^\alpha(w,n,m)$ & 0.399 & 0.597 & 0.034 & 0.041 & 0.039 & 0.280 \\
\hline
5 & 0,0,0,5,0 & $T(w,n,m)$        & 0.312 & 0.469 & 0.023 & 0.031 & 0.032 & 0.224 \\
  &  & $T^\alpha(w,n,m)$ & 0.365 & 0.549 & 0.038 & 0.039 & 0.045 & 0.274 \\
\hline
5 & 0,0,0,0,5 & $T(w,n,m)$        & 0.323 & 0.440 & 0.033 & 0.038 & 0.035 & 0.255 \\
  &  & $T^\alpha(w,n,m)$ & 0.370 & 0.516 & 0.044 & 0.046 & 0.047 & 0.305 \\
\hline
\hline
8 & 2,0,0,0,0,0,0,0 & $T(w,n,m)$        & 0.492 & 0.807 & 0.018 & 0.055 & 0.066 & 0.189 \\
  &  & $T^\alpha(w,n,m)$ & 0.538 & 0.843 & 0.024 & 0.067 & 0.086 & 0.229 \\
\hline
8 & 0,2,0,0,0,0,0,0 & $T(w,n,m)$        & 0.488 & 0.812 & 0.017 & 0.054 & 0.063 & 0.187 \\
  &  & $T^\alpha(w,n,m)$ & 0.572 & 0.863 & 0.022 & 0.065 & 0.073 & 0.225 \\
\hline
8 & 1,0,0,0,0,0,0,1 & $T(w,n,m)$        & 0.470 & 0.754 & 0.023 & 0.060 & 0.042 & 0.230 \\
  &  & $T^\alpha(w,n,m)$ & 0.526 & 0.803 & 0.031 & 0.075 & 0.053 & 0.275 \\
\hline
8 & 0,0,0,0,0,0,2,0 & $T(w,n,m)$        & 0.406 & 0.684 & 0.021 & 0.040 & 0.042 & 0.211 \\
  &  & $T^\alpha(w,n,m)$ & 0.457 & 0.730 & 0.027 & 0.050 & 0.053 & 0.241 \\
\hline
8 & 0,0,0,0,0,0,0,2 & $T(w,n,m)$        & 0.453 & 0.713 & 0.026 & 0.064 & 0.035 & 0.247 \\
  &  & $T^\alpha(w,n,m)$ & 0.511 & 0.768 & 0.036 & 0.080 & 0.048 & 0.310 \\
\hline
\hline
\end{tabular}}
\end{table}

\begin{table}
\label{t2}
\caption{ \scriptsize{ Power comparison for different hazard alternatives at 10\% significance level for several progressively censored samples with the sample size is $n=20$}}
\vspace{.05in}
\hspace{-0.5in}
{\begin{tabular}{|c|c|c||c|c||c|c||c|c|c}
\hline
\hline
 & & & \multicolumn{2}{|c||}{\small monotone increasing} & \multicolumn{2}{|c||}{\small monotone decreasing} & \multicolumn{2}{|c|}{\small nonmonotone} \\
 & & & \multicolumn{2}{|c||}{\small hazard alternatives} & \multicolumn{2}{|c||}{\small hazard alternatives} & \multicolumn{2}{|c|}{\small hazard alternatives} \\
\cline{1-9}
{m} & \small  schemes           & statistics &\small Gamma       &\small Weibull     &\small  Gamma       &\small Weibull     & \small  Beta       &\small Log-Normal     \\
    & \small $(R_1,\dots, R_m)$ &  &\small shape 2     &\small shape 2     &\small  shape 0.5   &\small  shape 0.5   & \small  shape 0.5     &\small  shape 1     \\
\hline
5 & 15,0,0,0,0 & $T(w,n,m)$        & 0.543 & 0.794 & 0.017 & 0.038 & 0.024 & 0.327 \\
  &  & $T^\alpha(w,n,m)$ & 0.600 & 0.844 & 0.022 & 0.048 & 0.036 & 0.394 \\
\hline
5 & 0,15,0,0,0 & $T(w,n,m)$        & 0.485 & 0.732 & 0.017 & 0.037 & 0.017 & 0.323 \\
  &  & $T^\alpha(w,n,m)$ & 0.533 & 0.792 & 0.024 & 0.053 & 0.022 & 0.357 \\
\hline
5 & 3,3,3,3,3 & $T(w,n,m)$        & 0.400 & 0.529 & 0.023 & 0.029 & 0.023 & 0.363 \\
  &  & $T^\alpha(w,n,m)$ & 0.483 & 0.590 & 0.030 & 0.041 & 0.030 & 0.417 \\
\hline
5 & 0,0,0,15,0 & $T(w,n,m)$        & 0.289 & 0.426 & 0.032 & 0.035 & 0.030 & 0.265 \\
  &  & $T^\alpha(w,n,m)$ & 0.337 & 0.479 & 0.043 & 0.048 & 0.045 & 0.311 \\
\hline
5 & 0,0,0,0,15 & $T(w,n,m)$        & 0.332 & 0.421 & 0.035 & 0.037 & 0.040 & 0.356 \\
  &  & $T^\alpha(w,n,m)$ & 0.376 & 0.475 & 0.044 & 0.044 & 0.044 & 0.400 \\
\hline
\hline
10 & 10,0,0,$\dots$,0,0,0 & $T(w,n,m)$        & 0.632 & 0.930 & 0.006 & 0.043 & 0.048 & 0.266 \\
  &  & $T^\alpha(w,n,m)$ & 0.675 & 0.935 & 0.060 & 0.226 & 0.094 & 0.360 \\
\hline
10 & 0,10,0,$\dots$,0,0,0 & $T(w,n,m)$        & 0.672 & 0.946 & 0.007 & 0.039 & 0.026 & 0.280 \\
  &  & $T^\alpha(w,n,m)$ & 0.703 & 0.948 & 0.061 & 0.231 & 0.065 & 0.374 \\
\hline
10 & 1,1,1,$\dots$,1,1,1 & $T(w,n,m)$        & 0.605 & 0.865 & 0.008 & 0.036 & 0.012 & 0.357 \\
  &  & $T^\alpha(w,n,m)$ & 0.630 & 0.875 & 0.080 & 0.195 & 0.060 & 0.443 \\
\hline
10 & 0,0,0,$\dots$,0,10,0 & $T(w,n,m)$        & 0.376 & 0.588 & 0.026 & 0.030 & 0.028 & 0.267 \\
  &  & $T^\alpha(w,n,m)$ & 0.404 & 0.623 & 0.079 & 0.126 & 0.075 & 0.301 \\
\hline
10 & 0,0,0,$\dots$,0,0,10 & $T(w,n,m)$        & 0.534 & 0.735 & 0.037 & 0.077 & 0.028 & 0.473 \\
  &  & $T^\alpha(w,n,m)$ & 0.584 & 0.779 & 0.117 & 0.197 & 0.085 & 0.522 \\
\hline
\hline
\end{tabular}}
\end{table}

\begin{table}
\label{t3}
\caption{ \scriptsize{ Power comparison for different hazard alternatives at 10\% significance level for several progressively censored samples with the sample size is $n=30$}}
\vspace{.05in}
\hspace{-0.5in}
{\begin{tabular}{|c|c|c||c|c||c|c||c|c|c}
\hline
\hline
 & & & \multicolumn{2}{|c||}{\small monotone increasing} & \multicolumn{2}{|c||}{\small monotone decreasing} & \multicolumn{2}{|c|}{\small nonmonotone} \\
 & & & \multicolumn{2}{|c||}{\small hazard alternatives} & \multicolumn{2}{|c||}{\small hazard alternatives} & \multicolumn{2}{|c|}{\small hazard alternatives} \\
\cline{1-9}
{m} & \small  schemes           & statistics &\small Gamma       &\small Weibull     &\small  Gamma       &\small Weibull     & \small  Beta       &\small Log-Normal     \\
    & \small $(R_1,\dots, R_m)$ &  &\small shape 2     &\small shape 2     &\small  shape 0.5   &\small  shape 0.5   & \small  shape 0.5     &\small  shape 1     \\
\hline
5 & 25,0,0,0,0 & $T(w,n,m)$        & 0.617 & 0.852 & 0.014 & 0.040 & 0.017 & 0.428 \\
  &  & $T^\alpha(w,n,m)$ & 0.678 & 0.885 & 0.022 & 0.051 & 0.025 & 0.490 \\
\hline
5 & 0,25,0,0,0 & $T(w,n,m)$        & 0.551 & 0.817 & 0.017 & 0.047 & 0.013 & 0.392 \\
  &  & $T^\alpha(w,n,m)$ & 0.611 & 0.851 & 0.026 & 0.056 & 0.018 & 0.447 \\
\hline
5 & 5,5,5,5,5 & $T(w,n,m)$        & 0.400 & 0.538 & 0.026 & 0.030 & 0.023 & 0.447 \\
  &  & $T^\alpha(w,n,m)$ & 0.456 & 0.589 & 0.032 & 0.034 & 0.029 & 0.492 \\
\hline
5 & 0,0,0,25,0 & $T(w,n,m)$        & 0.301 & 0.416 & 0.039 & 0.044 & 0.038 & 0.280 \\
  &  & $T^\alpha(w,n,m)$ & 0.356 & 0.502 & 0.054 & 0.063 & 0.052 & 0.340 \\
\hline
5 & 0,0,0,0,25 & $T(w,n,m)$        & 0.340 & 0.414 & 0.040 & 0.039 & 0.037 & 0.402 \\
  &  & $T^\alpha(w,n,m)$ & 0.405 & 0.491 & 0.055 & 0.051 & 0.051 & 0.481 \\
\hline
\hline
15 & 15,0,0,$\dots$,0,0,0 & $T(w,n,m)$        & 0.756 & 0.984 & 0.002 & 0.003 & 0.082 & 0.255 \\
  &  & $T^\alpha(w,n,m)$ & 0.760 & 0.974 & 0.237 & 0.599 & 0.270 & 0.459 \\
\hline
15 & 0,15,0,$\dots$,0,0,0 & $T(w,n,m)$        & 0.786 & 0.990 & 0.001 & 0.003 & 0.047 & 0.274 \\
  &  & $T^\alpha(w,n,m)$ & 0.784 & 0.982 & 0.250 & 0.609 & 0.243 & 0.471 \\
\hline
15 & 1,1,1,$\dots$,1,1,1 & $T(w,n,m)$        & 0.723 & 0.950 & 0.001 & 0.002 & 0.007 & 0.402 \\
  &  & $T^\alpha(w,n,m)$ & 0.780 & 0.965 & 0.341 & 0.588 & 0.249 & 0.597 \\
\hline
15 & 0,0,0,$\dots$,0,15,0 & $T(w,n,m)$        & 0.389 & 0.611 & 0.021 & 0.024 & 0.025 & 0.274 \\
  &  & $T^\alpha(w,n,m)$ & 0.425 & 0.664 & 0.179 & 0.307 & 0.175 & 0.323 \\
\hline
15 & 0,0,0,$\dots$,0,0,15 & $T(w,n,m)$        & 0.687 & 0.838 & 0.023 & 0.060 & 0.013 & 0.652 \\
  &  & $T^\alpha(w,n,m)$ & 0.735 & 0.907 & 0.343 & 0.517 & 0.286 & 0.696 \\
\hline
\hline
\end{tabular}}
\end{table}

\input{epsf}
\label{inc}
\epsfxsize=5in \epsfysize=5in
\begin{figure}
\hspace{0.25in}
\centerline{\epsfxsize=6in \epsfysize=3in \epsffile{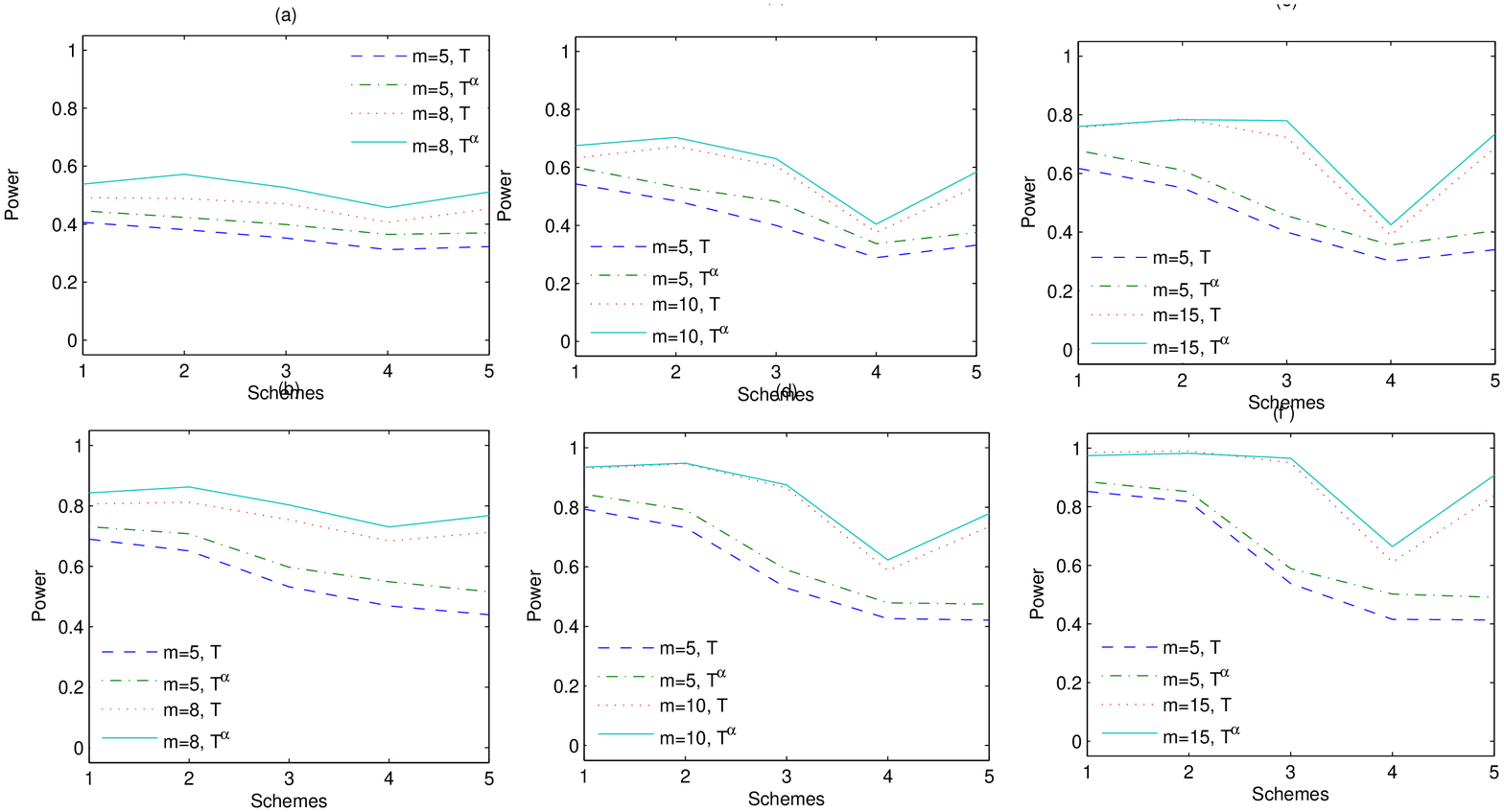}}
\vspace{-0.25in}
\centering\caption{\scriptsize Power comparison in monotone increasing hazard alternatives (Gamma: shape 2 (a, c, e), Weibull: shape 2 (b, d, f)) at 10\% significance level for several progressively censored samples when the sample size is 10 (a, b), 20 (c, d), and 30 (e, f). }
\end{figure}

\input{epsf}
\label{dec}
\epsfxsize=5in \epsfysize=5in
\begin{figure}
\hspace{0.05in}
\vspace{-0.1in}
\centerline{\epsfxsize=6.5in \epsfysize=3in \epsffile{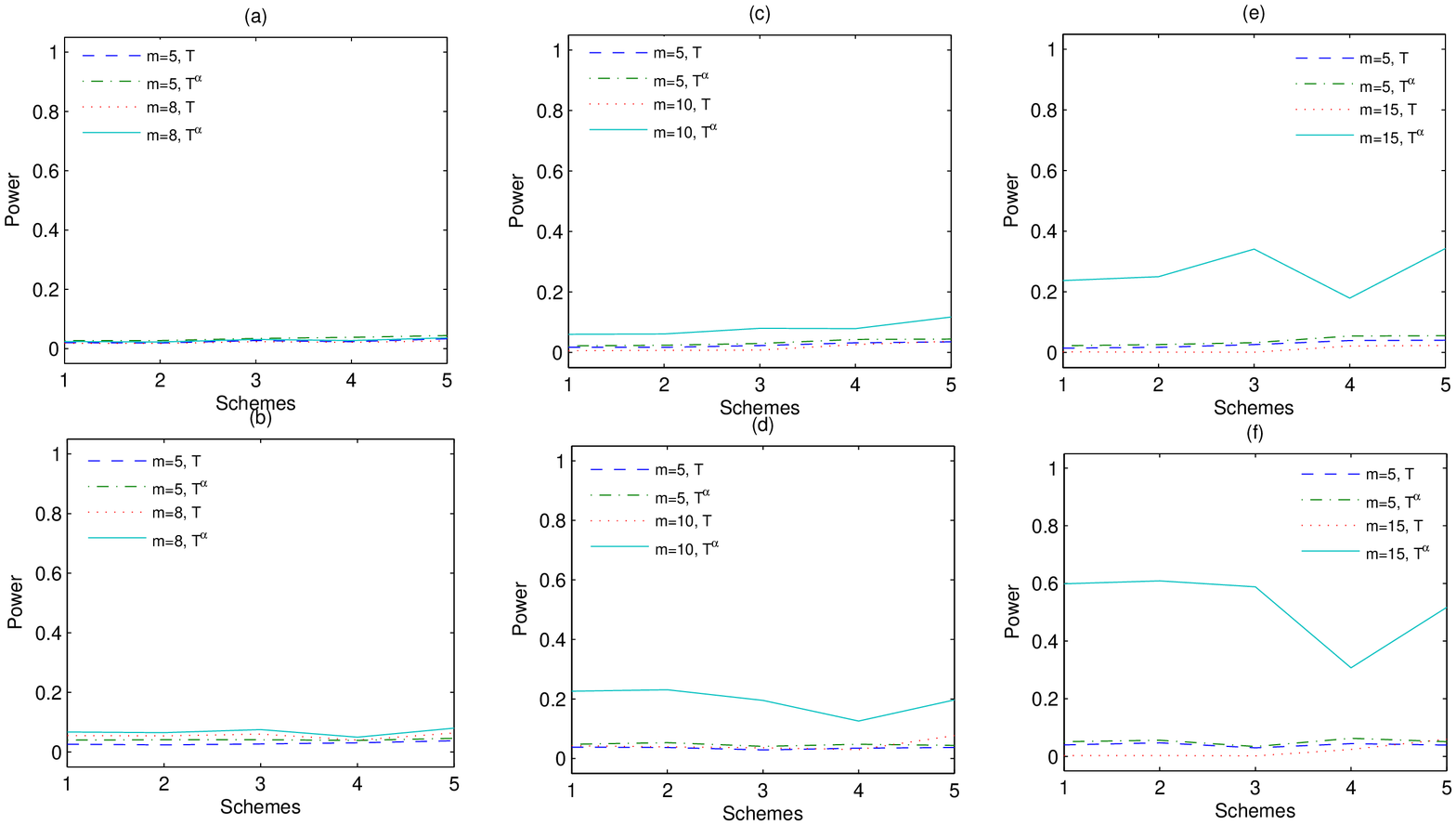}}
\vspace{-0.25in}
\centering\caption{\scriptsize Power comparison in monotone decreasing hazard alternatives (Gamma: shape 0.5 (a, c, e), Weibull: shape 0.5 (b, d, f)) at 10\% significance level for several progressively censored samples when the sample size is 10 (a, b), 20 (c, d), and 30 (e, f). }
\end{figure}

\input{epsf}
\label{non}
\epsfxsize=5in \epsfysize=5in
\begin{figure}
\hspace{0.25in}
\centerline{\epsfxsize=6in \epsfysize=3in \epsffile{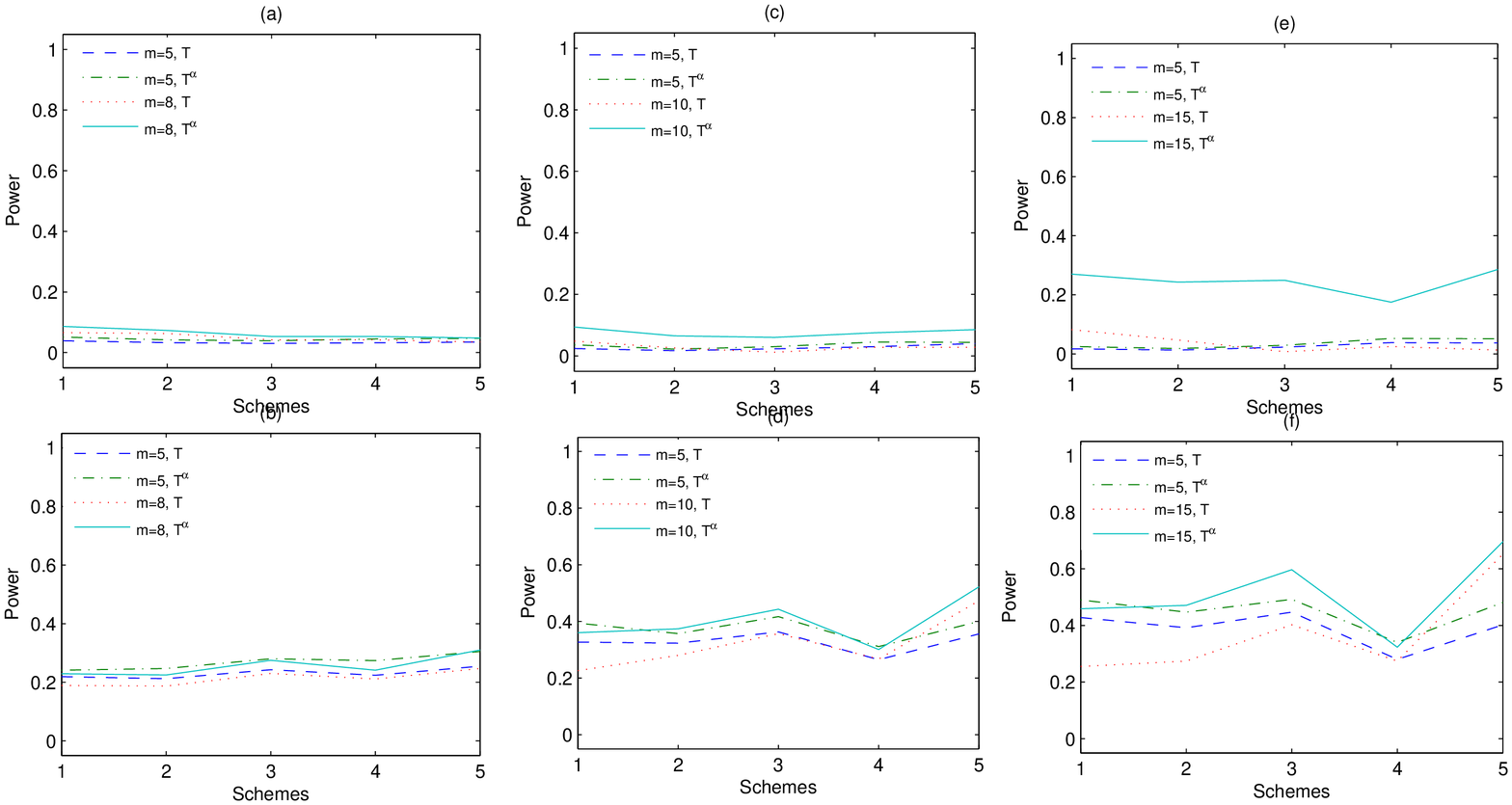}}
\vspace{-0.25in}
\centering\caption{\scriptsize Power comparison in nonmonotone hazard alternatives (Beta: shape 0.5 (a, c, e), Log-Normal: shape 1 (b, d, f)) at 10\% significance level for several progressively censored samples when the sample size is 10 (a, b), 20 (c, d), and 30 (e, f). }
\end{figure}

\section{Illustrative Example}
\par
In this section, we present an example to illustrate the use of the test statistic $T^\alpha(w,n,m)$ for testing the validity of the exponential distribution for an observed progressively Type-II right censored sample. We consider the data of Nelson \cite{12 Bala}, Table $6.1$, concerning times to breakdown of an insulating fluid tested at $34$ kilovolts. Viveros and Balakrishnan \cite{19 Bala} used Nelson's data, and randomly generated a progressively Type-II censored sample of size $m=8$ from $n=19$ observations. The data, and the progressive censoring scheme employed are given in Table 4.

\begin{table}
\label{t4}
\caption{{\scriptsize Nelson's data and progressively Type-II censoring scheme}}
\vspace{.05in}
\hspace{.75in}
{\begin{tabular}{|c||c|c|c|c|c|c|c|c|}
\hline
{\small\mbox{i}} & 1 & 2 & 3 & 4 & 5 & 6 & 7 & 8 \\
\hline
{\small \mbox{$x_{i:8:19}$}} & 0.19 & 0.78 & 0.96 & 1.31 & 2.78 & 4.85 & 6.5 & 7.35 \\
\hline
{\small \mbox{$R_i$}} & 0 & 0 & 3 & 0 & 3 & 0 & 0 & 5 \\
\hline
\end{tabular}}
\end{table}

\par
The test statistic computed from (\ref{8b}) is
\begin{equation*}
T^\alpha(w,n,m)=-\frac{1}{n}H^\alpha(w,n,m)+\frac{m}{n}\left[\log\left(\frac{1}{m}\sum_{j=1}^m(R_j+1)X_{j:m:n}\right)+1\right]=0.2422,
\end{equation*}
where $w=3$ according to Table I in \cite{15 Bala}. We estimate the unknown $\theta$ by the maximum likelihood estimator, and so the null distribution can be approximated by an exponential distribution with $\hat{\theta}=9.09$. The $p$-value is then computed as
\begin{equation*}
P\left(T^\alpha(w,n,m)>0.2422|H_0:f_{X}(x)=\frac{1}{9.09}\exp\left(\frac{-x}{9.09}\right)\right)=0.9737
\end{equation*}
which provides very strong evidence that the observed progressively Type-II censored sample is from an exponential distribution.

\section{Conclusion}
\par
In R\'enyi entropy, with controlling $\alpha$ about 1, we provided a test statistic which was quite powerful when compared to an existing goodness-of-fit test proposed for progressively Type-II censored data due to \cite{3 My paper}. Because the sampling distributions of these test statistics were intractable, we determined the percentage points using $10,000$ Monte Carlo samples from an exponential distribution.
\par
We saw from the Figures 1-3 and Tables 1-3 that the proposed test statistic showed better powers than the competing test statistics against the alternatives with monotone increasing, monotone decreasing and nonmonotone hazard functions, which apply to many real-life applications. Also, we saw from Figures 1-3 and Tables 1-3 that the scheme $(R_1=n-m, R_2=0, \dots, R_m=0)$ showed higher power than the other schemes when the alternative was a monotone increasing hazard function. For the alternative with monotone decreasing hazard functions, the scheme $(R_1=0, \dots, R_{m-1}=0, R_{m}=n-m)$ (the conventional Type-II censored data) showed higher power. Finally, for the alternative with a nonmonotone hazard function, sometimes the former censoring scheme gave a higher power, and sometimes the latter censoring scheme did.
\par
This work has the potential to be applied in the context of censored data and goodness of fit tests. This paper can elaborate further researches by extending such modifications for other censoring schemes. Also, this area of research can be expanded by considering other distributions besides the exponential distribution such as Pareto, Log-Normal and Weibull distributions. Finally, the method of $\alpha$ determination can be developed by considering an optimality algorithm which maximize the power.

\begin{center}
{\bf Appendix A}
\end{center}
{\em Proof of Theorem 2.1}
\par
By the decomposition property of the entropy measure in \cite{Pasha}, we have
\begin{equation*}
H^\alpha_{1\cdots m:m:n}=H^\alpha_{1:m:n}+H^\alpha_{2|1:m:n}+\cdots+H^\alpha_{r|r-1:m:n}+\cdots+H^\alpha_{m|m-1:m:n}.
\end{equation*}
The conditional p.d.f of progressively Type-II censored order statistics is given in \cite{13 Bala} as
\begin{equation*}
f_{X_{r:m:n}|X_{r-1:m:n}}(x_r|x_{r-1})
=\left(n-\sum_{j=1}^{r-1}R_j-r+1\right)h(x_r)\left[\frac{1-F(x_r)}{1-F(x_{r-1})}
\right]^{n-\sum_{j=1}^{r-1}R_j-r+1},
\end{equation*}
where $h(x_r)=\frac{f(x_r)}{1-F(x_{r-1})}$ and so $f_{X_{r:m:n}|X_{r-1:m:n}}(x_r|x_{r-1})$ can be interpreted as the density of the first order statistic among an $n-\sum_{j=1}^{r-1}R_j-r+1$ sample from $\frac{f(x)}{1-F(x_{r-1})}$, so that we can obtain by (\ref{4b})
\begin{equation*}
\hspace{-1.5in}H^\alpha_{r|r-1:m:n}=-\log\left(n-\sum_{j=1}^{r-1}R_j-r+1\right)
\end{equation*}
\begin{equation*}
\hspace{1in}+\frac{1}{1-\alpha}\log\int_{-\infty}^{\infty}f_{X_{r:m:n}}(x)\{h(x)\}^{\alpha-1}\left\{1-F_{X_{r:m:n}}(x)\right\}^{\alpha-1}\mathrm{d}x.
\end{equation*}
The required result then follows.
\\
\begin{center}
{\bf Appendix B}
\end{center}
{\em Proof of Lemma 2.1}
\par
The p.d.f of $X_{r:m:n}$ is given in \cite{3 Bala} as
\begin{equation*}
f_{X_{j:m:n}}(x)=c_{j-1}f(x)\sum_{i=1}^ja_{i,j}(1-F(x))^{\gamma_i-1},
\end{equation*}
where $\gamma_i=m-i+1+\sum_{j=i}^mR_j$ for $1\leq i\leq m$, $c_{j-1}=\prod_{u=1}^j\gamma_u$ for $1\leq j\leq m$, $a_{i,j}=\prod_{u=1}^j1/(\gamma_u-\gamma_i)$ for $1\leq i\leq j\leq m$, and $\gamma_u-\gamma_i=1$ for $u=i$. So $\bar{H}^{\alpha}_{1\cdots m:m:n}$ is equal to
\begin{eqnarray*}
&&\frac{1}{1-\alpha}\sum_{j=1}^{m}\log\int_{-\infty}^{\infty}f_{X_{j:m:n}}(x)\{h(x)\}^{\alpha-1}\left\{1-F_{X_{j:m:n}}(x)\right\}^{\alpha-1}\mathrm{d}x\\
&=&\frac{1}{1-\alpha}\sum_{j=1}^m\log\int_{-\infty}^{\infty}c_{j-1}f(x)\sum_{i=1}^ja_{i,j}\{1-F(x)\}^{\gamma_i-1}\{h(x)\}^{\alpha-1}
\left\{1-F_{X_{j:m:n}}(x)\right\}^{\alpha-1}\mathrm{d}x\\
&=&\frac{1}{1-\alpha}\sum_{j=1}^m\log E\left(\int_{-\infty}^{F^{-1}(U_{1:\gamma_i-1})}c_{j-1}f(x)\sum_{i=1}^ja_{i,j}\{h(x)\}^{\alpha-1}
\left\{1-F_{X_{j:m:n}}(x)\right\}^{\alpha-1}\mathrm{d}x\right)\\
&=&\frac{1}{1-\alpha}\sum_{j=1}^m\log E\left(\int_{-\infty}^{F^{-1}(U_{1:\gamma_i-1})}c_{j-1}f(x)\sum_{i=1}^ja_{i,j}\frac{\{f^{-1}(x)\}^{1-\alpha}}{\{1-F(x)\}^{\alpha-1}}\right.
\end{eqnarray*}
\begin{equation*}
\hspace{1in}\times\left.\left[\sum_{u=0}^{j-1}\binom{n}{u}\{F(x)\}^u\{1-F(x)\}^{n-u}\right]^{\alpha-1}\mathrm{d}x\right)\qquad \left[\mbox{Put:}\;\; F(x)=p\right]
\end{equation*}
\begin{equation*}
\hspace{.22in}=\frac{1}{1-\alpha}\sum_{j=1}^m\log E\left(\int_{-\infty}^{U_{1:\gamma_i-1}}c_{j-1}\sum_{i=1}^ja_{i,j}\frac{\{\frac{\mathrm{d}F^{-1}(p)}{\mathrm{d}p}\}^{1-\alpha}}{(1-p)^{\alpha-1}}\right.
\left.\left[\sum_{u=0}^{j-1}\binom{n}{u}p^u(1-p)^{n-u}\right]^{\alpha-1}\mathrm{d}p\right).
\end{equation*}
Hence, the lemma is proved.

\bibliographystyle{plain}

\end{document}